\documentstyle[epsfig]{article}                                                         
\font\tenbf=cmbx10                                                              
\font\tenrm=cmr10                                                               
\font\tenit=cmti10                                                              
\font\elevenbf=cmbx10 scaled\magstep 1                                          
\font\elevenrm=cmr10 scaled\magstep 1                                           
\font\elevenit=cmti10 scaled\magstep 1                                          
                                                              
\font\ninerm=cmr9

\newcommand{\ci}[2]{${}^{\mbox{\tenrm #2}}$}
\newcommand{\ice}[1]{\relax}
  \newcommand{\EQN}{\label}                                                     
  \newcommand{\beq}{\begin{equation}}                                           
  \newcommand{\eeq}{\end{equation}}                                             
  \newcommand{\r}[1]{Eq.~(\ref{#1})}                                                

\newcommand{\al}{\alpha}                                                        
\newcommand{\be}{\beta}                                                         
\newcommand{\ep}{\epsilon}                                                      
\newcommand{\bea}{\begin{eqnarray}}                                             
\newcommand{\eea}{\end{eqnarray}}

\newcommand{\ba}{\begin{array}}                                                 
\newcommand{\ea}{\end{array}}

\newcommand{\G}{\Gamma}                                                         
\newcommand{\g}{\gamma}

\newcommand{\dsp}{\displaystyle}                                                
                                                    
\def\bbuildrel#1_#2^#3%
{\mathrel{\mathop{\kern 0pt#1}\limits_{#2}^{#3}}}

\newcommand{\prd}{\partial}                                                     
                                                                                
\newcommand{\myfrac}[2]{\frac{\dsp #1}{\dsp #2}}                                
                                                                                
\renewenvironment{thebibliography}[1]                                           
 { \elevenrm                                                                    
   \begin{list}{\arabic{enumi}.}                                                
    {\usecounter{enumi} \setlength{\parsep}{0pt}                                
     \setlength{\itemsep}{3pt} \settowidth{\labelwidth}{#1.}                    
     \sloppy                                                                    
    }}{\end{list}}                                                              
                                                                                
\begin{document}                                                                
\begin{center}{{\tenbf A CLOSED ANALYTICAL FORMULA \\                           
               \vglue 8pt                                                      
               FOR  TWO-LOOP MASSIVE TADPOLES WITH                              
               \\                                                               
               \vglue 3pt                                                       
                ARBITRARY TENSOR NUMERATORS\footnote{Published in {\small
in: {\em New Computing Techniques in Physics Research III},
eds. K.-H. Becks and D. Perret-Gallix
(World Scientific, Singapore, 1994),
p.~559.}}
               \\}                                                              
\vglue 1.0cm                                                                    
{\tenrm K.G. CHETYRKIN \\}                                                       
\baselineskip=13pt                                                              
{\tenit Institute for Nuclear Research of the Russian Academy of Sciences,      
\\}                                                                             
\baselineskip=12pt                                                              
{\tenit Moscow, 117312, Russia\\}                                               
\vglue 0.8cm                                                                    
{\tenrm ABSTRACT}}                                                              
\end{center}                                                                    
\vglue 0.3cm                                                                    
{\rightskip=3pc                                                                 
 \leftskip=3pc                                                                  
 \tenrm\baselineskip=12pt                                                       
 \noindent                                                                      
Using the integration by parts method we derive a closed analytical 
expression for the result of the integration of an arbitrary 
dimensionally regulated tadpole diagram composed of  a massless 
propagator   and two massive ones, each  raised into an arbitrary 
power, and including an arbitrary tensor numerator.  We also briefly 
discuss the implementation of the  formula  in the algebraic 
manipulation language of FORM.                                                  
\vglue 0.6cm}                                                                   
{\elevenbf\noindent 1. Introduction}                                            
\baselineskip=14pt                                                              
\elevenrm                                                                       
\vglue 0.4cm                                                                    
Feynman integrals (FI) with complicated tensor numerators are usually                
difficult to work with. Even  in the cases when  they are known to be           
analytically calculable {\em in principle}, in practical terms their            
evaluation often implies a tedious, time-consuming and error-prone              
labour of reducing the problem to the calculation of a host of                  
properly constructed scalar integrals.  Moreover, the number  of the            
latter integrals grows very fast  as the numerator's structure gets             
more complicated.                                                               
                                                                                
There exist only a few examples when the task is completely solved.  
A good example is provided with the so-called {\em $p$-integrals}, 
that is completely massless Feynman integrals depending on only one 
external momentum. Here the explicit result  is known  for one-loop 
integrals\ci{me81}{1}, while, say, a given three-loop  tensor 
$p$-integral can be done  by employing  a rather cumbersome and 
time-costly method of harmonic projections\ci{LarTkaVer91}{2}.

In this talk we discuss another useful class of Feynman integrals --- 
integrals without external momenta at all but comprising  massive 
lines\footnote{\ninerm It is understood that all the massive 
propagators depend on one and the  same mass $m$.} as well as  
massless ones.  They will be referred to as {\em  $m$-integrals}.  
Such integrals naturally appear in  many problems where the mass $m$ 
may be treated as  a "heavy" one, much larger than all other mass scales 
involved.

In the one-loop case $m$-integrals are rather trivial and  we shall 
concentrate on   two-loop $m$-integrals pictured in Fig. 1.  2-loop 
$m$-integrals with only one massive line (Fig. 1a )
may be reduced to 1-loop 
case after firstly integrating the 1-loop p-subintegral.  2-loop 
$m$-integrals with more than 1 massive lines (Fig. 1b,c) are not  
so easy to do.  We  show  how the use of the integration by parts 
method leads to a simple and general result for arbitrary (not 
necessarily scalar) two-loop $m$-integral with two massive and one 
massless line (see Fig. 1b). In principle, our method allows also to 
reduce a tensor integral with 3 massive lines of Fig. 1c                    
to a similar scalar integral. The latter can probably be done 
(at least in some particular cases) again  through integration by 
parts\ci{Dav}{3}.  But to the best of our knowledge no explicit 
integration formula  for this integral exists if the powers of all three 
propagators are arbitrary.

{
\begin {figure}                                        
\begin{center}
\begin {tabular}{ccc}  
\parbox{3cm}{   
\epsfig{file=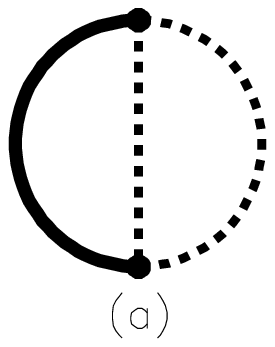,width=3.cm,height=3.cm} 
            }
&
\parbox{3cm}{
\epsfig{file=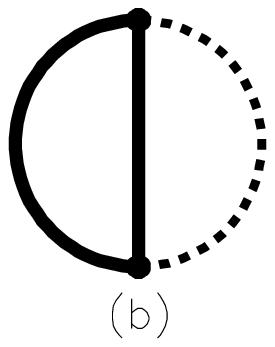,width=3.cm,height=3.cm}
            }
&
\parbox{3cm}{ 
\epsfig{file=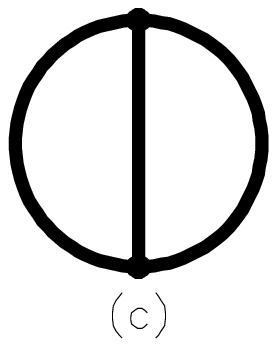,width=3.cm,height=3.cm} 
            }  
\end {tabular}      
\end{center}
\caption {
\tenrm
Diffferent cases of 2-loop $m$-integrals:
dashed lines are massless; solid lines have mass $m$.
}
\end{figure}   
%
\vglue 0.6cm                                                                    
{\elevenbf\noindent 2. Setting the problem}                                     
\vglue 0.4cm                                                                    
We begin with a bit more complicated FI of the same topology as shown           
in Fig. 1b but with a  non-zero external momentum $q$.          
The corresponding analytical  expression reads              
(in Euclidean space)
\beq                                                                            
\myfrac{1}{(\pi^2)^{2-\ep}}                                                     
\int \myfrac{ {\rm d}^D \ell_1\, {\rm d}^D \ell_2\,\, {\cal P}(p)}%
{(p_1^2 + m^2)^\al                                                              
(p_2^2 +m^2)^\be (p_3^2)^{\g}}                                                 
\EQN{2loop.tadpole}                                                             
\eeq                                                                            
where $\ell_1$ and $\ell_2$ are the loop momenta, $p =                        
\{p_1,p_2,p_3\}$ are the propagator momenta    and ${\cal P}()$ is a            
tensor  nominator. We shall deal with  three  possibilities                     
of expressing the momenta $p$ in terms of the loop momenta and the            
external  momentum, viz.,                                                   
\begin{eqnarray}                                                                
&& p_1 = \ell_1 + q, \ \ p_2 = \ell_2, \ \ p_3 = -(\ell_1 + \ell_2),            
\label{mom1}                                                                       
\\                                                                              
&&                                                                              
p_1 = \ell_1 , \ \ p_2 = \ell_2 + q, \ \ p_3 = -(\ell_1 + \ell_2),              
\label{mom2}                                                                       
\\                                                                              
&&                                                                              
p_1 = \ell_1, \ \ p_2 = \ell_2, \ \ p_3 = q -(\ell_1 + \ell_2).                 
\label{mom3}                                                                       
\end{eqnarray}

In a particular case of   ${\cal P}()\equiv 1, \ \,  q=0$                       
the result for \r{2loop.tadpole} is known\ci{Veltman.2loopTadpole}{4}           
\[                                                                              
\mbox{(\rm \protect\r{2loop.tadpole} with ${\cal P}(p)\equiv 1$,            
and }   q=0 )  =                                                                
(m^2)^{D- \al - \be - \g }                                                       
\myfrac{\G(D/2  -\gamma)}{\G(\al)\G(\be)\G(D/2)} M(\al,\be,\gamma)              
\]                                                                              
with                                                                            
\beq                                                                            
M(\al,\be,\gamma)                                                               
=                                                                               
\myfrac{%
\G(\al + \g - {D}/{2})                                                         
\G(\be + \g - {D}/{2})                                                          
\G(\al + \be + \g - D)}                                                         
{\G(\al + \be + 2\g - D)
}
.                                                       
\EQN{2loop.res}                                                                 
\eeq                                                                            
                                                                                
Our aim is to generalize this result  on the  case                              
of arbitrary tensor polynomial ${\cal P}(p)$, still 
keeping $q = 0$.  To simplify the formulas we
shall set $m=1$  below.
\vglue 0.6cm                                                                    
{\elevenbf\noindent 2.                                                          
Solution through the integration by parts  method}                              
\vglue 0.4cm                                                                    
As is well known within dimensional regularization the value of the             
FI \r{2loop.tadpole} does not depend on specifying the propagator momenta      
and all  three choices  (\ref{mom1}-\ref{mom3}) are completely equivalent.            
The fact can be conveniently expressed through some differential                 
identities.  The essence of the integration by part method for                  
dimensionally regulated Feynman integrals  consists of the                      
use of such identities  in order to simplify integrals                          
to be computed\ci{me81}{1}.                                                     
                                                                                
Let us try to apply the method in our case.                                     
It is convenient to write eq. \r{2loop.tadpole} 
in a condensed form as follows           
\beq                                                                            
\mbox{eq. \protect\r{2loop.tadpole}}                                            
=                                                                               
\int                                                                            
{\cal P} (p) I(\al,\be,\g)
\EQN{abb}                                                                       
\eeq                                                                            
and to introduce  four differential operators acting on \r{abb}                 
$$ \tilde{p}_i^\mu  = \myfrac{1}{2}\myfrac{\prd}{\prd p_i^\mu},                 
\ \                                                                             
i=1,2,3 \ \ \mbox{{\rm and}}  \ \                                               
\tilde{q}^\mu  = \myfrac{1}{2}\myfrac{\prd}{\prd q^\mu}.                        
$$                                                                              
Explicitly, one has                                                             
$$                                                                              
\int I(\al,\be,\g) = 
\myfrac{\G(D/2  -\gamma)}{\G(\al)\G(\be)\G(D/2)}
M(\al,\be,\g),                                             
$$                                                                              
\beq                                                                            
\ba{c}                                                                          
\tilde{p}_i^\mu                                                                 
\left(                                                                          
{\cal P} (p) I(\al,\be,\g)                                                  
\right)                                                                         
=                                                                               
\left(                                                                          
\myfrac{1}{2}\myfrac{\prd}{\prd p_i^\mu}                                        
{\cal P} (p)                                                                
\right)                                                                         
I(\al,\be,\g)                                                                   
-                                                                               
\alpha_i                                                                        
{\cal P} (p)                                                                
I(\al + \delta_{1i},\be+ \delta_{2i},\g + \delta_{3i})
{p}_i^\mu,
\\                                                                              
\tilde{q}^\mu                                                                   
\left(                                                                          
{\cal P} (p) I(\al,\be,\g)                                                  
\right)                                                                         
=                                                                               
\left(                                                                          
\myfrac{1}{2}\myfrac{\prd}{\prd p_3^\mu}                                        
{\cal P} (p)                                                                
\right)                                                                         
I(\al,\be,\g)                                                                   
-                                                                               
\gamma                                                                          
{\cal P} (p)                                                                
I(\al,\be,\g + 1)
p_3^\mu,     
\ea                                                                             
\EQN{oper.defs}                                                                 
\eeq                                                                            
with $\alpha_1 =\al, \alpha_2 = \be,                                            
\alpha_3 = \g$.                                                                 
The equivalence of the three momentum patterns                                      
(\ref{mom1}-\ref{mom3}) may now be expressed as  a chain of                           
identities                                                                      
\beq                                                                            
\int \tilde{p}_1^\mu  {\cal P} (p) I(\al,\be,\g)                            
=\int \tilde{p}_2^\mu {\cal P} (p) I(\al,\be,\g)                            
=\int \tilde{p}_3^\mu  {\cal P} (p) I(\al,\be,\g)                           
=\int\tilde{q}^\mu  {\cal P} (p) I(\al,\be,\g)  .                               
\EQN{chain}                                                                     
\eeq                                                                            
                                                                                
These equalities  allow  us to                                                  
evaluate immediately  an integral of the form 
\beq                                                                            
\int \tilde{{\cal P}}_{2n} (\tilde{p}) I(\al,\be,\g)                             
|_{{}_{\scriptstyle q=0}}                                                   
=                                                                               
\myfrac{\G(D/2  -\gamma)}{\G(\al)\G(\be)\G(D/2)}
\left\{
\left(\myfrac{\Box_{\tilde{q}}}{4}\right)^n                                     
\tilde{{\cal P}}( \tilde{q})_{2n}%
\right\}
\myfrac{(-)^n (\gamma)_n  }{n!(2 - \ep)_n}                                 
M(\al,\be,\gamma +n),                                                                     
\EQN{res1}                                                                      
\eeq                                                                            
with the Pochhammer symbol                                                      
$(a)_n = \myfrac{\Gamma(a+ n )}{\Gamma(a)}$,
$D=4 - 2\ep$                                      
and $\tilde{{\cal P}}_{2n}(\tilde{p})$ being an arbitrary                         
tensor in $\tilde{p}= \{\tilde{p}_1,\tilde{p}_2,\tilde{p}_3\}$
of rank $2n$.
Indeed, in view of \r{chain} we may freely replace                               
$\tilde{{\cal P}}_{2n}(\tilde{p})$ by                                            
$\tilde{{\cal P}}_{2n}(\tilde{q}) \equiv                                          
\tilde{{\cal P}}_{2n}(\tilde{p_1}=\tilde{q},\tilde{p_2}=\tilde{q},                 
\tilde{p_3}=\tilde{q}).                                                         
$                                                                               
Now the result \r{res1} comes from                                              
three simple observations:                                                      
                                                                                
(i)                                                                             
The integral $\int I(\al,\be,\g)$ is  a scalar function                         
of $q$ and thus only the scalar component of the polynomial                   
$\tilde{{\cal P}}_{2n}(\tilde{q})$ (that is proportional to                        
$\tilde{q}^{2n}$) will survive  after setting                                            
$\tilde{q}=0$ in the very end of the calculation.                                       
                                                                                
(ii)                                                                            
$\left(\myfrac{\Box_{\tilde{q}}}{4}\right)^n                                    
(\tilde{q})^n = n!(2 - \ep)_n                                                  
$.                                                                              
                                                                                
(iii)                                                                            
$\left(\myfrac{\Box_{\tilde{q}}}{4}\right)^n                               
\myfrac{1}{\tilde{q}^{2\gamma}} = (\gamma)_n 
(\gamma - 1 + \ep)_n                    
\myfrac{1}{(\tilde{q}^2)^{\gamma + n}}
= (-)^n (\gamma)_n (2 -  \ep - \gamma -n )_n 
\myfrac{1}{(\tilde{q}^2)^{\gamma + n}}
$.                                                                              
                                                                                
Thus  we are left with the task of finding a representation of the 
initial integral \r{2loop.tadpole} as a linear combinations of 
integrals of the form  displayed in \r{res1}. The problem is solved 
by the use of the following   identity                                                        
\beq                                                                            
{\cal P}_n(q) f(q^2)  = \sum_{\sigma=0}^{[n/2]}                                 
\myfrac{(-)^\sigma}{4^\sigma\sigma!}                                                         
\left\{                                                                         
\Box^\sigma_{\tilde{q}} {\cal P}_n(\tilde{q})                                   
\right\}                                                                        
f^{(-n + \sigma)}(q^2) 
\EQN{idi}                                                                      
\eeq                                                                            
where  $f(x)$  is an arbitrary smooth function of x 
and $f^{(n)}(x)$ is defined in such a way that                                  
\beq                                                                            
\frac{d}{d x} f^{(n)}(x)  \equiv                                            
f^{(n+1)}(x)                                                                    
\EQN{def}                                                                       
.                                                                               
\eeq                                                                            
It should be clear now that once the initial                                           
integral \r{2loop.tadpole} has been                                                      
expressed as a linear combinations of integrals of the form \r{res1} it         
may be done without any problem.                                                 
Indeed, without essential loss of generality we may assume
that the polynomial   ${\cal P}(p) = {\cal P}_{n_1,n_2} (p)$  
does not depend on
$p_3$ and  meets the following homogeneity equation:
\[
{\cal P}_{n_1,n_2} (\lambda_1 p_1,\lambda_2 p_2)
\equiv \lambda_1^{n_1} \lambda_2^{n_2}
{\cal P}_{n_1,n_2} (p_1, p_2). 
\]
Now a direct application of  \r{chain}, \r{res1} and \r{idi} gives
\beq                                                                            
\ba{c}                                                                          
\dsp  
\int {{\cal P}}_{n_1,n_2} (p_1,p_2) I(\al,\be,\g)                         
=
\\                                                                              
\dsp                                                                            
\myfrac{\G(D/2 - \gamma)  }{\G(\al)\G(\be)\G(D/2)}               
\sum_{\sigma_3=0}^{[(n_1 + n_2)/2]}                                             
\sum_{\sigma_2=0}^{[n_2/2]}                                                     
\sum_{\sigma_1=0}^{[n_1/2]}                                                     
\myfrac{(\g)_{\sigma_3}}{\sigma_3 ! (2 - \ep)_{\sigma_3}}                       
M(\al + \sigma_1 -n_1,\be + \sigma_2 -n_2,\g + \sigma_3)                        
\\                                                                              
\dsp                                                                            
\myfrac{(-)^{(n_1 + n_2 + \sigma_3)}}                                                      
{4^{(\sigma_1 + \sigma_2 + \sigma_3)}\sigma_1 !\sigma_2 !}                                                     
\left\{                                                                         
\Box^{\sigma_3}_{p_3}                                                           
\left\{                                                                         
\Box^{\sigma_2}_{p_2}                                                           
\Box^{\sigma_1}_{p_1}                                                           
{\cal P}_{n_1,n_2}(p_1,p_2)                                                    
\right\}|_{{}_{\scriptstyle p_1 = p_2 = p_3}}                                   
\right\}|_{{}_{\scriptstyle p_3=0}}                                             
\ea
\EQN{res3}                                                                      
\eeq                                                                            
which is the formula we wanted.                                                                                

The algebraic structure of \r{res3} is very similar to that of
the corresponding formula for 1-loop $p$-integrals in\ci{me81}{1}.
This observation has allowed us  to perform  a simple
algebraic programming of \r{res3} in FORM\ci{Ver91}{5} 
by closely following the routine    ONE.PRC
from the package MINCER\ci{LarTkaVer91}{2}.
\vglue 0.5cm                                                                    
{\elevenbf \noindent 5. Acknowledgments \hfil}                                 
\vglue 0.4cm                                                                    
I appreciate the warm hospitality of Institute of Theoretical
Particle Physics at the Karlsruhe University where 
this work  has been finished. 
I thank   Mrs. M. Frasure and A. Kwiatkowski 
for their help in preparing the manuscript.
A stipendium from 
the HERAUES-stiftung is gratefully acknowledged.
\vglue 0.5cm                                                                    
{\elevenbf\noindent 6. References \hfil}                                        
\vglue 0.4cm                                                                    
                                                           
\end{document}